\documentclass[preprint,prd,tightenlines,superscriptaddress]{revtex4-1}
\usepackage{graphicx} % Include figure files
\usepackage{dcolumn}  % Align table columns on decimal point
\usepackage{colordvi}
\usepackage{color}
\usepackage{epstopdf}
\usepackage{amssymb}
\usepackage{url}
\graphicspath{{ps}}
\usepackage{hyperref}
\usepackage{tabularx}
\usepackage{float} % manually added for [H]
\usepackage{bm}% manually added for mathbf
\usepackage{subfigure} %to use subfigure
\usepackage{comment} %to comment out
 \usepackage[compat=1.0.0]{tikz-feynman}
\usepackage{tikz-feynman}
\usepackage{hyperref}
\usepackage{tasks}

\usepackage{placeins}

\usepackage[english]{babel}

\input belle2sym.tex

\begin{document}

%place for definitions and newcommands
\def\belletwo {\it {Belle II}}

\vspace*{-3\baselineskip}
\resizebox{!}{3cm}{\includegraphics{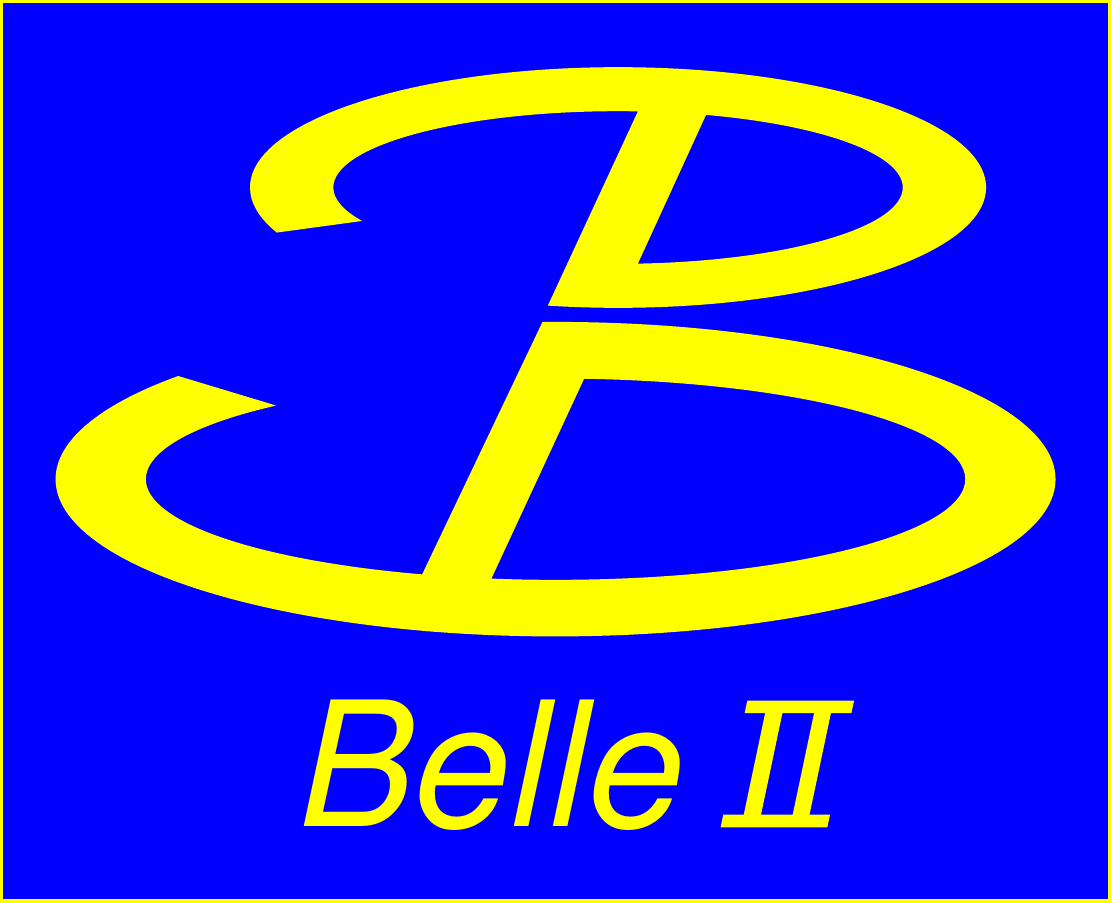}}

\vspace*{-5\baselineskip}
\begin{flushright}
BELLE2-CONF-PH-2021-014\\
%Version 8.0 \\
%\today
\end{flushright}

\title { \quad\\[0.5cm] Measurements of the branching fractions for {\boldmath $B \to K^{*}\gamma$} decays at Belle II}

%%% Paper:    (2021 conference papers)
%%% Journal:  (2021 conferences)
%%% Updates:
%%% July 21, 2021 - new iteration
%%% ====================================================================
%%% Use \input{authors-conf2021} to insert this material into your latex file.
%%%\newcommand{\instSinica}{Academia Sinica, Taipei 11529, Taiwan}
\newcommand{\instCPPM}{Aix Marseille Universit\'{e}, CNRS/IN2P3, CPPM, 13288 Marseille, France}
\newcommand{\instYerevan}{Alikhanyan National Science Laboratory, Yerevan 0036, Armenia}
\newcommand{\instBeihang}{Beihang University, Beijing 100191, China}
\newcommand{\instBNL}{Brookhaven National Laboratory, Upton, New York 11973, U.S.A.}
\newcommand{\instBINP}{Budker Institute of Nuclear Physics SB RAS, Novosibirsk 630090, Russian Federation}
\newcommand{\instCMU}{Carnegie Mellon University, Pittsburgh, Pennsylvania 15213, U.S.A.}
\newcommand{\instCinvestavIPN}{Centro de Investigacion y de Estudios Avanzados del Instituto Politecnico Nacional, Mexico City 07360, Mexico}
\newcommand{\instPrague}{Faculty of Mathematics and Physics, Charles University, 121 16 Prague, Czech Republic}
\newcommand{\instChiangMai}{Chiang Mai University, Chiang Mai 50202, Thailand}
\newcommand{\instChiba}{Chiba University, Chiba 263-8522, Japan}
\newcommand{\instChonnam}{Chonnam National University, Gwangju 61186, South Korea}
\newcommand{\instConacyt}{Consejo Nacional de Ciencia y Tecnolog\'{\i}a, Mexico City 03940, Mexico}
\newcommand{\instDESY}{Deutsches Elektronen--Synchrotron, 22607 Hamburg, Germany}
\newcommand{\instDuke}{Duke University, Durham, North Carolina 27708, U.S.A.}
\newcommand{\instITAR}{Institute of Theoretical and Applied Research (ITAR), Duy Tan University, Hanoi 100000, Vietnam}
\newcommand{\instRomaENEA}{ENEA Casaccia, I-00123 Roma, Italy}
\newcommand{\instFuJen}{Department of Physics, Fu Jen Catholic University, Taipei 24205, Taiwan}
\newcommand{\instFudan}{Key Laboratory of Nuclear Physics and Ion-beam Application (MOE) and Institute of Modern Physics, Fudan University, Shanghai 200443, China}
\newcommand{\instGoettingen}{II. Physikalisches Institut, Georg-August-Universit\"{a}t G\"{o}ttingen, 37073 G\"{o}ttingen, Germany}
\newcommand{\instGifu}{Gifu University, Gifu 501-1193, Japan}
\newcommand{\instSOKENDAI}{The Graduate University for Advanced Studies (SOKENDAI), Hayama 240-0193, Japan}
\newcommand{\instGyeongsang}{Gyeongsang National University, Jinju 52828, South Korea}
\newcommand{\instHanyang}{Department of Physics and Institute of Natural Sciences, Hanyang University, Seoul 04763, South Korea}
\newcommand{\instKEK}{High Energy Accelerator Research Organization (KEK), Tsukuba 305-0801, Japan}
\newcommand{\instJPARC}{J-PARC Branch, KEK Theory Center, High Energy Accelerator Research Organization (KEK), Tsukuba 305-0801, Japan}
\newcommand{\instHiroshima}{Hiroshima University, Higashi-Hiroshima, Hiroshima 739-8530, Japan}
\newcommand{\instFrascati}{INFN Laboratori Nazionali di Frascati, I-00044 Frascati, Italy}
\newcommand{\instNapoliINFN}{INFN Sezione di Napoli, I-80126 Napoli, Italy}
\newcommand{\instPadovaINFN}{INFN Sezione di Padova, I-35131 Padova, Italy}
\newcommand{\instPerugiaINFN}{INFN Sezione di Perugia, I-06123 Perugia, Italy}
\newcommand{\instPisaINFN}{INFN Sezione di Pisa, I-56127 Pisa, Italy}
\newcommand{\instRomaINFN}{INFN Sezione di Roma, I-00185 Roma, Italy}
\newcommand{\instRomaTreINFN}{INFN Sezione di Roma Tre, I-00146 Roma, Italy}
\newcommand{\instTorinoINFN}{INFN Sezione di Torino, I-10125 Torino, Italy}
\newcommand{\instTriesteINFN}{INFN Sezione di Trieste, I-34127 Trieste, Italy}
\newcommand{\instIISER}{Indian Institute of Science Education and Research Mohali, SAS Nagar, 140306, India}
\newcommand{\instIITBhubaneswar}{Indian Institute of Technology Bhubaneswar, Satya Nagar 751007, India}
\newcommand{\instIITGuwahati}{Indian Institute of Technology Guwahati, Assam 781039, India}
\newcommand{\instIITHyderabad}{Indian Institute of Technology Hyderabad, Telangana 502285, India}
\newcommand{\instIITMadras}{Indian Institute of Technology Madras, Chennai 600036, India}
\newcommand{\instIndiana}{Indiana University, Bloomington, Indiana 47408, U.S.A.}
\newcommand{\instIHEPRussia}{Institute for High Energy Physics, Protvino 142281, Russian Federation}
\newcommand{\instHEPHYVienna}{Institute of High Energy Physics, Vienna 1050, Austria}
\newcommand{\instIHEPChina}{Institute of High Energy Physics, Chinese Academy of Sciences, Beijing 100049, China}
\newcommand{\instIPP}{Institute of Particle Physics (Canada), Victoria, British Columbia V8W 2Y2, Canada}
\newcommand{\instIOP}{Institute of Physics, Vietnam Academy of Science and Technology (VAST), Hanoi, Vietnam}
\newcommand{\instIFIC}{Instituto de Fisica Corpuscular, Paterna 46980, Spain}
\newcommand{\instISU}{Iowa State University, Ames, Iowa 50011, U.S.A.}
\newcommand{\instJAEA}{Advanced Science Research Center, Japan Atomic Energy Agency, Naka 319-1195, Japan}
\newcommand{\instMainz}{Institut f\"{u}r Kernphysik, Johannes Gutenberg-Universit\"{a}t Mainz, D-55099 Mainz, Germany}
\newcommand{\instGiessen}{Justus-Liebig-Universit\"{a}t Gie\ss{}en, 35392 Gie\ss{}en, Germany}
\newcommand{\instKarlsruhe}{Institut f\"{u}r Experimentelle Teilchenphysik, Karlsruher Institut f\"{u}r Technologie, 76131 Karlsruhe, Germany}
\newcommand{\instKitasato}{Kitasato University, Sagamihara 252-0373, Japan}
\newcommand{\instKISTI}{Korea Institute of Science and Technology Information, Daejeon 34141, South Korea}
\newcommand{\instKoreaUnivKU}{Korea University, Seoul 02841, South Korea}
\newcommand{\instKSU}{Kyoto Sangyo University, Kyoto 603-8555, Japan}
\newcommand{\instKyungpook}{Kyungpook National University, Daegu 41566, South Korea}
\newcommand{\instLPI}{P.N. Lebedev Physical Institute of the Russian Academy of Sciences, Moscow 119991, Russian Federation}
\newcommand{\instLNNU}{Liaoning Normal University, Dalian 116029, China}
\newcommand{\instLMU}{Ludwig Maximilians University, 80539 Munich, Germany}
\newcommand{\instLuther}{Luther College, Decorah, Iowa 52101, U.S.A.}
\newcommand{\instMNITJaipur}{Malaviya National Institute of Technology Jaipur, Jaipur 302017, India}
\newcommand{\instMPP}{Max-Planck-Institut f\"{u}r Physik, 80805 M\"{u}nchen, Germany}
\newcommand{\instMPGHLL}{Semiconductor Laboratory of the Max Planck Society, 81739 M\"{u}nchen, Germany}
\newcommand{\instMcGill}{McGill University, Montr\'{e}al, Qu\'{e}bec, H3A 2T8, Canada}
\newcommand{\instMEPhI}{Moscow Physical Engineering Institute, Moscow 115409, Russian Federation}
\newcommand{\instNagoya}{Graduate School of Science, Nagoya University, Nagoya 464-8602, Japan}
\newcommand{\instNagoyaIAR}{Institute for Advanced Research, Nagoya University, Nagoya 464-8602, Japan}
\newcommand{\instNagoyaKMI}{Kobayashi-Maskawa Institute, Nagoya University, Nagoya 464-8602, Japan}
\newcommand{\instNaraWu}{Nara Women's University, Nara 630-8506, Japan}
\newcommand{\instHSE}{National Research University Higher School of Economics, Moscow 101000, Russian Federation}
\newcommand{\instNTUTaiwan}{Department of Physics, National Taiwan University, Taipei 10617, Taiwan}
\newcommand{\instNUUTaiwan}{National United University, Miao Li 36003, Taiwan}
\newcommand{\instKrakow}{H. Niewodniczanski Institute of Nuclear Physics, Krakow 31-342, Poland}
\newcommand{\instNiigata}{Niigata University, Niigata 950-2181, Japan}
\newcommand{\instNSU}{Novosibirsk State University, Novosibirsk 630090, Russian Federation}
\newcommand{\instOkinawa}{Okinawa Institute of Science and Technology, Okinawa 904-0495, Japan}
\newcommand{\instOsakaCity}{Osaka City University, Osaka 558-8585, Japan}
\newcommand{\instRCNP}{Research Center for Nuclear Physics, Osaka University, Osaka 567-0047, Japan}
\newcommand{\instPNNL}{Pacific Northwest National Laboratory, Richland, Washington 99352, U.S.A.}
\newcommand{\instPanjab}{Panjab University, Chandigarh 160014, India}
\newcommand{\instPanjabPAU}{Punjab Agricultural University, Ludhiana 141004, India}
\newcommand{\instRIKENMSL}{Meson Science Laboratory, Cluster for Pioneering Research, RIKEN, Saitama 351-0198, Japan}
\newcommand{\instXavier}{St. Francis Xavier University, Antigonish, Nova Scotia, B2G 2W5, Canada}
\newcommand{\instSeoul}{Seoul National University, Seoul 08826, South Korea}
\newcommand{\instSPU}{Showa Pharmaceutical University, Tokyo 194-8543, Japan}
\newcommand{\instSoochow}{Soochow University, Suzhou 215006, China}
\newcommand{\instSoongsil}{Soongsil University, Seoul 06978, South Korea}
\newcommand{\instLjubljanaJSI}{J. Stefan Institute, 1000 Ljubljana, Slovenia}
\newcommand{\instKyiv}{Taras Shevchenko National Univ. of Kiev, Kiev, Ukraine}
\newcommand{\instTata}{Tata Institute of Fundamental Research, Mumbai 400005, India}
\newcommand{\instTUM}{Department of Physics, Technische Universit\"{a}t M\"{u}nchen, 85748 Garching, Germany}
\newcommand{\instTelAviv}{Tel Aviv University, School of Physics and Astronomy, Tel Aviv, 69978, Israel}
\newcommand{\instToho}{Toho University, Funabashi 274-8510, Japan}
\newcommand{\instTohoku}{Department of Physics, Tohoku University, Sendai 980-8578, Japan}
\newcommand{\instTitech}{Tokyo Institute of Technology, Tokyo 152-8550, Japan}
\newcommand{\instTokyoMetropolitan}{Tokyo Metropolitan University, Tokyo 192-0397, Japan}
\newcommand{\instUAS}{Universidad Autonoma de Sinaloa, Sinaloa 80000, Mexico}
\newcommand{\instNapoliUNIV}{Dipartimento di Scienze Fisiche, Universit\`{a} di Napoli Federico II, I-80126 Napoli, Italy}
\newcommand{\instPadovaUNIV}{Dipartimento di Fisica e Astronomia, Universit\`{a} di Padova, I-35131 Padova, Italy}
\newcommand{\instPerugiaUNIV}{Dipartimento di Fisica, Universit\`{a} di Perugia, I-06123 Perugia, Italy}
\newcommand{\instPisaUNIV}{Dipartimento di Fisica, Universit\`{a} di Pisa, I-56127 Pisa, Italy}
\newcommand{\instRomaTreUNIV}{Dipartimento di Matematica e Fisica, Universit\`{a} di Roma Tre, I-00146 Roma, Italy}
\newcommand{\instTorinoUNIV}{Dipartimento di Fisica, Universit\`{a} di Torino, I-10125 Torino, Italy}
\newcommand{\instTriesteUNIV}{Dipartimento di Fisica, Universit\`{a} di Trieste, I-34127 Trieste, Italy}
\newcommand{\instMontreal}{Universit\'{e} de Montr\'{e}al, Physique des Particules, Montr\'{e}al, Qu\'{e}bec, H3C 3J7, Canada}
\newcommand{\instIJCLab}{Universit\'{e} Paris-Saclay, CNRS/IN2P3, IJCLab, 91405 Orsay, France}
\newcommand{\instIPHC}{Universit\'{e} de Strasbourg, CNRS, IPHC, UMR 7178, 67037 Strasbourg, France}
\newcommand{\instAdelaide}{Department of Physics, University of Adelaide, Adelaide, South Australia 5005, Australia}
\newcommand{\instBonn}{University of Bonn, 53115 Bonn, Germany}
\newcommand{\instUBC}{University of British Columbia, Vancouver, British Columbia, V6T 1Z1, Canada}
\newcommand{\instCincinnati}{University of Cincinnati, Cincinnati, Ohio 45221, U.S.A.}
\newcommand{\instFlorida}{University of Florida, Gainesville, Florida 32611, U.S.A.}
\newcommand{\instHawaii}{University of Hawaii, Honolulu, Hawaii 96822, U.S.A.}
\newcommand{\instHeidelberg}{University of Heidelberg, 68131 Mannheim, Germany}
\newcommand{\instLjubljanaUniLJ}{Faculty of Mathematics and Physics, University of Ljubljana, 1000 Ljubljana, Slovenia}
\newcommand{\instLouisville}{University of Louisville, Louisville, Kentucky 40292, U.S.A.}
\newcommand{\instMalaya}{National Centre for Particle Physics, University Malaya, 50603 Kuala Lumpur, Malaysia}
\newcommand{\instLjubljanaUM}{Faculty of Chemistry and Chemical Engineering, University of Maribor, 2000 Maribor, Slovenia}
\newcommand{\instMelbourne}{School of Physics, University of Melbourne, Victoria 3010, Australia}
\newcommand{\instMississippi}{University of Mississippi, University, Mississippi 38677, U.S.A.}
\newcommand{\instUOM}{University of Miyazaki, Miyazaki 889-2192, Japan}
\newcommand{\instPittsburgh}{University of Pittsburgh, Pittsburgh, Pennsylvania 15260, U.S.A.}
\newcommand{\instUSTC}{University of Science and Technology of China, Hefei 230026, China}
\newcommand{\instSAlabama}{University of South Alabama, Mobile, Alabama 36688, U.S.A.}
\newcommand{\instSCarolina}{University of South Carolina, Columbia, South Carolina 29208, U.S.A.}
\newcommand{\instSydney}{School of Physics, University of Sydney, New South Wales 2006, Australia}
\newcommand{\instUTokyo}{Department of Physics, University of Tokyo, Tokyo 113-0033, Japan}
\newcommand{\instEri}{Earthquake Research Institute, University of Tokyo, Tokyo 113-0032, Japan}
\newcommand{\instIPMU}{Kavli Institute for the Physics and Mathematics of the Universe (WPI), University of Tokyo, Kashiwa 277-8583, Japan}
\newcommand{\instVictoria}{University of Victoria, Victoria, British Columbia, V8W 3P6, Canada}
\newcommand{\instVPI}{Virginia Polytechnic Institute and State University, Blacksburg, Virginia 24061, U.S.A.}
\newcommand{\instWayneState}{Wayne State University, Detroit, Michigan 48202, U.S.A.}
\newcommand{\instYamagata}{Yamagata University, Yamagata 990-8560, Japan}
\newcommand{\instYonsei}{Yonsei University, Seoul 03722, South Korea}
%%%\newcommand{\instZZU}{Zhengzhou University, Zhengzhou 450001, China}
%%%\affiliation{\instSinica}
\affiliation{\instCPPM}
\affiliation{\instYerevan}
\affiliation{\instBeihang}
%%%\affiliation{\instBUAP}
\affiliation{\instBNL}
\affiliation{\instBINP}
\affiliation{\instCMU}
\affiliation{\instCinvestavIPN}
\affiliation{\instPrague}
\affiliation{\instChiangMai}
\affiliation{\instChiba}
\affiliation{\instChonnam}
%%%\affiliation{\instChula}
\affiliation{\instConacyt}
\affiliation{\instDESY}
\affiliation{\instDuke}
\affiliation{\instITAR}
\affiliation{\instRomaENEA}
%%%\affiliation{\instJuelich}
\affiliation{\instFuJen}
\affiliation{\instFudan}
\affiliation{\instGoettingen}
\affiliation{\instGifu}
\affiliation{\instSOKENDAI}
\affiliation{\instGyeongsang}
\affiliation{\instHanyang}
\affiliation{\instKEK}
\affiliation{\instJPARC}
\affiliation{\instHiroshima}
%%%\affiliation{\instHUNNU}
\affiliation{\instFrascati}
\affiliation{\instNapoliINFN}
\affiliation{\instPadovaINFN}
\affiliation{\instPerugiaINFN}
\affiliation{\instPisaINFN}
\affiliation{\instRomaINFN}
\affiliation{\instRomaTreINFN}
\affiliation{\instTorinoINFN}
\affiliation{\instTriesteINFN}
\affiliation{\instIISER}
\affiliation{\instIITBhubaneswar}
\affiliation{\instIITGuwahati}
\affiliation{\instIITHyderabad}
\affiliation{\instIITMadras}
\affiliation{\instIndiana}
\affiliation{\instIHEPRussia}
\affiliation{\instHEPHYVienna}
\affiliation{\instIHEPChina}
%%%\affiliation{\instChennai}
\affiliation{\instIPP}
\affiliation{\instIOP}
\affiliation{\instIFIC}
\affiliation{\instISU}
\affiliation{\instJAEA}
\affiliation{\instMainz}
\affiliation{\instGiessen}
\affiliation{\instKarlsruhe}
%%%\affiliation{\instKennesaw}
\affiliation{\instKitasato}
\affiliation{\instKISTI}
\affiliation{\instKoreaUnivKU}
\affiliation{\instKSU}
%%%\affiliation{\instKyotoU}
\affiliation{\instKyungpook}
\affiliation{\instLPI}
\affiliation{\instLNNU}
\affiliation{\instLMU}
\affiliation{\instLuther}
\affiliation{\instMNITJaipur}
\affiliation{\instMPP}
\affiliation{\instMPGHLL}
\affiliation{\instMcGill}
%%%\affiliation{\instMETU}
\affiliation{\instMEPhI}
\affiliation{\instNagoya}
\affiliation{\instNagoyaIAR}
\affiliation{\instNagoyaKMI}
%%%\affiliation{\instNNU}
\affiliation{\instNaraWu}
%%%\affiliation{\instUNAM}
\affiliation{\instHSE}
\affiliation{\instNTUTaiwan}
\affiliation{\instNUUTaiwan}
\affiliation{\instKrakow}
\affiliation{\instNiigata}
\affiliation{\instNSU}
\affiliation{\instOkinawa}
\affiliation{\instOsakaCity}
\affiliation{\instRCNP}
\affiliation{\instPNNL}
\affiliation{\instPanjab}
%%%\affiliation{\instPeking}
\affiliation{\instPanjabPAU}
\affiliation{\instRIKENMSL}
%%%\affiliation{\instRIKEN}
\affiliation{\instXavier}
\affiliation{\instSeoul}
%%%\affiliation{\instShandong}
\affiliation{\instSPU}
\affiliation{\instSoochow}
\affiliation{\instSoongsil}
\affiliation{\instLjubljanaJSI}
\affiliation{\instKyiv}
\affiliation{\instTata}
\affiliation{\instTUM}
%%%\affiliation{\instECUTUM}
\affiliation{\instTelAviv}
\affiliation{\instToho}
\affiliation{\instTohoku}
\affiliation{\instTitech}
\affiliation{\instTokyoMetropolitan}
\affiliation{\instUAS}
%%%\affiliation{\instNapoliUNIVA}
\affiliation{\instNapoliUNIV}
\affiliation{\instPadovaUNIV}
\affiliation{\instPerugiaUNIV}
\affiliation{\instPisaUNIV}
%%%\affiliation{\instRomaUNIV}
\affiliation{\instRomaTreUNIV}
\affiliation{\instTorinoUNIV}
\affiliation{\instTriesteUNIV}
\affiliation{\instMontreal}
\affiliation{\instIJCLab}
\affiliation{\instIPHC}
\affiliation{\instAdelaide}
\affiliation{\instBonn}
\affiliation{\instUBC}
\affiliation{\instCincinnati}
\affiliation{\instFlorida}
%%%\affiliation{\instHamburg}
\affiliation{\instHawaii}
\affiliation{\instHeidelberg}
\affiliation{\instLjubljanaUniLJ}
\affiliation{\instLouisville}
\affiliation{\instMalaya}
\affiliation{\instLjubljanaUM}
\affiliation{\instMelbourne}
\affiliation{\instMississippi}
\affiliation{\instUOM}
%%%\affiliation{\instNovaGorica}
\affiliation{\instPittsburgh}
\affiliation{\instUSTC}
\affiliation{\instSAlabama}
\affiliation{\instSCarolina}
\affiliation{\instSydney}
%%%\affiliation{\instTabuk}
\affiliation{\instUTokyo}
\affiliation{\instEri}
\affiliation{\instIPMU}
\affiliation{\instVictoria}
\affiliation{\instVPI}
\affiliation{\instWayneState}
\affiliation{\instYamagata}
\affiliation{\instYonsei}
%%%\affiliation{\instZZU}
  \author{F.~Abudin{\'e}n}\affiliation{\instTriesteINFN} % 2250
  \author{I.~Adachi}\affiliation{\instKEK}\affiliation{\instSOKENDAI} % 2590
  \author{R.~Adak}\affiliation{\instFudan} % 6743
  \author{K.~Adamczyk}\affiliation{\instKrakow} % 2239
  \author{L.~Aggarwal}\affiliation{\instPanjab} % 10083
  \author{P.~Ahlburg}\affiliation{\instBonn} % 2367
  \author{H.~Ahmed}\affiliation{\instXavier} % 11323
  \author{J.~K.~Ahn}\affiliation{\instKoreaUnivKU} % 7423
  \author{H.~Aihara}\affiliation{\instUTokyo} % 2223
  \author{N.~Akopov}\affiliation{\instYerevan} % 9443
% ...
% 4983
\collaboration{Belle II Collaboration}

\begin{abstract}
This paper reports a study of $B \to K^{*}\gamma$ decays using $62.8\pm 0.6$ fb$^{-1}$ of data collected during 2019--2020 by the Belle II experiment at the SuperKEKB $e^{+}e^{-}$ asymmetric-energy collider, corresponding to $(68.2 \pm 0.8) \times 10^6$ $B\overline{B}$ events. We find $454 \pm 28$, $50 \pm 10$, $169 \pm 18$, and $160 \pm 17$ signal events in the decay modes $B^{0} \to K^{*0}[K^{+}\pi^{-}]\gamma$, $B^{0} \to K^{*0}[K^0_{\rm S}\pi^{0}]\gamma$,  $B^{+} \to K^{*+}[K^{+}\pi^{0}]\gamma$, and $B^{+} \to K^{*+}[K^{+}\pi^{0}]\gamma$, respectively. The uncertainties quoted for the signal yield are statistical only. We report the branching fractions of these decays:
$$\mathcal{B} [B^{0} \to K^{*0}[K^{+}\pi^{-}]\gamma] = (4.5 \pm 0.3 \pm 0.2) \times 10^{-5}, $$
$$\mathcal{B} [B^{0} \to K^{*0}[K^0_{\rm S}\pi^{0}]\gamma] = (4.4 \pm 0.9 \pm 0.6) \times 10^{-5},$$
$$\mathcal{B} [B^{+} \to K^{*+}[K^{+}\pi^{0}]\gamma] = (5.0 \pm 0.5 \pm 0.4)\times 10^{-5},\text{ and}$$
$$\mathcal{B} [B^{+} \to K^{*+}[K^0_{\rm S}\pi^{+}]\gamma] = (5.4 \pm 0.6 \pm 0.4) \times 10^{-5},$$ where the first uncertainty is statistical, and the second is systematic. The results are consistent with world-average values.

\keywords{Belle II, ...}
\end{abstract}

\pacs{}

\maketitle

{\renewcommand{\thefootnote}{\fnsymbol{footnote}}}
\setcounter{footnote}{0}

%\tableofcontents
\pagebreak

\section{Introduction}
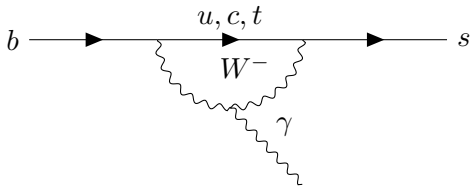
\begin{figure}[htb]
\begin{tikzpicture}
\begin{feynman}
\vertex (b1) { $b$};
\vertex [right=5em of b1] (b2);
\vertex [right=5em of b2] (b3);
\vertex [right=5em of b3] (b4) { $s$};

\vertex at ($(b2)!0.5!(b3)!0.9cm!-90:(b3)$) (g1);
\vertex [below=5em of b3] (g2);

\diagram* {

	(b1) -- [fermion] (b2) -- [fermion, edge label={$u,c,t$}] (b3) -- [fermion] (b4),
	(b2) -- [boson, quarter right] (g1) -- [boson, edge label = {{$W^{-}$}}, quarter right] (b3),
	(g1) -- [photon, edge label={ $\gamma $}] (g2),
};

\end{feynman}
\end{tikzpicture}
\caption{Leading order $b \to s \gamma$ loop diagram.}
\label{FD}
\end{figure}

The radiative decay $B\to K^{*}(892) \gamma$ is a flavor-changing neutral current process, which is forbidden at tree level in the standard model (SM) of particle physics. The transition proceeds dominantly through a one-loop $b \to s \gamma$ diagram. The contribution from annihilation diagrams is highly suppressed by factors of $\mathcal{O}(\lambda_{\text{QCD}}/m_{b})$ and Cabibbo-Kobayashi-Maskawa matrix elements~\cite{CKM}, where $\lambda_{\text{QCD}}$ is the location of Landau pole of quantum chromodynamics and $m_{b}$ is the mass of the $b$ quark. The largest SM contribution to the $b \to s \gamma$ transition is from the diagram shown in Fig.~\ref{FD} having a $t$ quark and $W$ boson in the loop. Throughout this document, $K^{*}$ implies a $K^{*}(892)$ meson and charge conjugate processes are included implicitly unless stated otherwise.

Extensions of the SM predict new particles that can contribute to the loop, potentially altering the branching fraction as well as other observables from their SM predictions, making the decay an excellent probe for such models~\cite{BSM1,BSM2}. These observables include the $CP$ violation asymmetry
	$$A_{CP} = \frac{\Gamma(\overline{B}\to\overline{K}^{*}\gamma) - \Gamma(B\to K^{*}\gamma)}{\Gamma(\overline{B}\to\overline{K}^{*}\gamma) + \Gamma(B\to K^{*}\gamma)} $$ and the isospin asymmetry $$
	\Delta_{0+} = \frac{\Gamma(B^{0}\to K^{*0}\gamma) - \Gamma(B^{+}\to K^{*+}\gamma)}{\Gamma(B^{0}\to K^{*0}\gamma) + \Gamma(B^{+}\to K^{*+}\gamma)}.$$

The SM prediction of the branching fraction suffers from large uncertainties related to form factors~\cite{SM1,SM2}. In contrast, observables like $ A_{CP}$ and $\Delta_{0+}$ are theoretically clean due to cancellation of these factors in the ratio~\cite{SM3,SM4}. The latest measurement by the Belle experiment~\cite{Belle_paper} with 771$\times10^{6}$ $B\overline{B}$ pairs, reported the first evidence for isospin violation at $3.1\sigma$ significance. Earlier to that, the CLEO~\cite{CLEO_paper} and BaBar~\cite{BaBar_paper} Collaborations had also performed similar measurements.  This brief summary of the current status of experimental and theoretical studies demonstrates that the $B \to K^{*}\gamma$ channel provides an ideal ground for indirect searches for new physics effects and tests of SM predictions. The current study presents preliminary results of branching fractions of $K^{*}\gamma$ modes measured using $e^{+}e^{-}$ collision data collected in the period of 2019--2020 by the Belle II detector. The measurement of observables like $A_{CP}$ and $\Delta_{0+}$ will be done when Belle II accumulates a data sample equivalent to that used in the Belle study in order to have similar sensitivities. 

\section{The Belle II Detector and Dataset}
Belle II is a large-solid-angle magnetic spectrometer designed to study products of $e^{+}e^{-}$ collisions. The detector is located at the collision point of the SuperKEKB accelerator~\cite{superkekb}. It is composed of several components arranged in a cylindrical geometry around the beam pipe. The innermost region of the detector comprises two subdetectors, namely two layers of DEPFET-based silicon pixel detector (PXD) and four layers of double-sided silicon strip detectors (SVD). The combination of PXD and SVD constitutes the inner tracking system. The measurement of charge and momentum of charged particle tracks is facilitated by a 56-layered central drift chamber, which also helps in particle identification (PID) by measuring the specific-ionization information. A Cherenkov-light angle and time-of-propagation detector situated in the barrel region and a proximity-focusing aerogel ring-imaging Cherenkov counter placed in the forward region together constitute the key PID system. An electromagnetic calorimeter (ECL) consisting of CsI(Tl) crystals measures the energy of photons and assists with electron identification. These subdetectors are located inside a superconducting solenoid coil that provides a 1.5\,T magnetic field. The return yoke of the magnet is instrumented with plastic scintillators and resistive plate chambers to identify $K^{0}_{\rm L}$ mesons and muons. Further details about the detector can be found in Ref.~\cite{belle2tdr}. 

The data sample used in this analysis was collected by Belle II in the period of 2019--2020 at a center-of-mass (CM) energy corresponding to the mass of the $\Upsilon(4S)$ resonance. The integrated luminosity is $62.8 \pm 0.6$ fb$^{-1}$, which is equivalent to $(68.2 \pm 0.8)\times 10^6$ $B \overline{B}$ events. The method used to compute the number of $B \overline{B}$ events is documented in Ref.~\cite{nbb}, but uses an updated sample of data and Monte Carlo (MC). 
To study the properties of signal events, optimize selection criteria, and determine detection efficiency, two million signal MC events are generated for all four $B \to K^{*}\gamma$ modes. In addition, inclusive $B\overline{B}$ and $q\overline{q}$ continuum MC samples are used for background classification, where $q$ denotes $u, d, s, \text{ and } c$ quark. The sample size is equivalent to an integrated luminosity of $3~\mathrm{ab}^{-1}$. These events are generated using the EvtGen~\cite{{EVETGEN}} package. Geant4~\cite{GEANT4} is used to simulate detector response. The Belle II analysis software framework~\cite{basf2} is used to process data.
Systematic uncertainties are studied using $9.2 \pm 0.9$ fb$^{-1}$ of off-resonance data and 500 fb$^{-1}$ of off-resonance MC events. The off-resonance data were collected at 60 MeV below the $\Upsilon(4S)$ resonance.

\section{Event selection and Reconstruction}

Photons, charged kaons and pions are reconstructed and identified using information from ECL, PID and tracking systems. We require the distance of closest approach to the interaction point in the plane transverse to the beam axis ($x$-$y$ plane) ${|d_{0}|< 2.0~\rm cm}$ and along the beam axis ($z$ axis) ${|d_{z}| < 4.0~\rm cm}$ to select charged tracks that originate from a region near the $e^{+}e^{-}$ collision point. A charged track is identified as a $K^{\pm}$ or $\pi^{\pm}$ using a likelihood ratio ${\mathcal{P}(K/\pi)=\frac{L_{K}}{L_{K}+L_{\pi}}}$, where ${L_{K}}$ and ${L_{\pi}}$ are the likelihood for a track to be a kaon or a pion, calculated based on inputs from PID subdetectors. We apply a criterion ${\mathcal{P}(K/\pi)> 0.6}$ to select ${K^{\pm}}$ and ${\mathcal{P}(\pi/K)> 0.6}$ to select ${\pi^{\pm}}$ candidates. 

As the process $B\to K^{*}\gamma$ is a two-body decay, in the $B$ rest frame we expect the prompt $\gamma$ candidate to have an energy of around half the $B$ meson mass ($\approx 2.5$ GeV$/c^{2}$). High-energy photon candidates are selected from the barrel region of the ECL and required to have an energy $2.25 <  E^{*}_{\gamma} < 2.85$ GeV. Here, and elsewhere in this paper, the superscript $^*$ implies that the quantity is calculated in the CM frame. To ensure the selection of an isolated high-energy photon, we apply selection criteria on variables based on the ECL shower shape and CsI(Tl) pulse shape discrimination information of the candidate~\cite{PSDMVA}. Photons coming from decays of $\pi^{0}$ or $\eta$ mesons constitute a major background for the analysis. We reject such photon candidates having kinematics consistent with that of $\pi^{0}$ or $\eta$ decay product. 

The $K^0_{\rm S}$ candidates are reconstructed from a pair of oppositely charged tracks, assumed to be pions, and kinematically fit assuming they originate from a common vertex. Candidates that fail the vertex fit are rejected. Selections on $|d_{0}|$ and $|z_{0}|$, as well as PID criteria are not applied to these tracks. We further apply requirements on the kinematic variables of $K^0_{\rm S}$ candidates namely: momentum-dependent criteria on the $K^0_{\rm S}$ flight length in the transverse plane, azimuthal angle between the momentum vector and the vector between the interaction point and the decay vertex of the $K^0_{\rm S}$ candidate, and the distance at the interaction point along the $z$ axis of the two tracks used to reconstruct the $K^0_{\rm S}$ candidate. The invariant-mass window for $K^0_{\rm S}$ candidates is 0.488 to 0.508 ${\rm GeV}\!/c^2$, which corresponds to about $\pm 6 \sigma$ around the nominal $K^0_{\rm S}$ mass. Here, $\sigma$ is the resolution obtained by fitting the invariant mass distribution of correctly reconstructed signal candidates. We apply such relaxed criteria to incorporate non-Gaussian tails in the mass distribution. 

The signal side $\pi^{0}$ is reconstructed from a pair of photons each having an energy greater than 80, 30, or 60 MeV, depending on whether the photon is detected in the forward, barrel, or backward region of ECL, respectively. The invariant mass window for $\pi^{0}$ candidates is $ 0.120 \text{ to } 0.145 \,{\rm GeV}\!/c^2$. We define the helicity angle for $\pi^{0}$ as the angle between the line defined by the momentum difference of the photons from $\pi^{0}$ calculated in the $\pi^{0}$ frame and the momentum of $\pi^{0}$ candidate calculated in the lab frame. We place a requirement on the helicity angle of the $\pi^{0}$ candidate to suppress combinatorial background. 

The $K^{*}$ candidate is reconstructed by combining a kaon ($K^{\pm}$ or $K^0_{\rm S}$) with a pion ($\pi^{\pm}$ or $\pi^{0}$). We retain $K^{*}$ candidates inside the invariant-mass window $ 0.817 \text{ to }0.967 \,{\rm GeV}\!/c^2$, which corresponds to around three times the natural width of the $K^{*}$ meson. As $K^{*}$ is a vector meson decaying to a pair of pseudoscalar mesons, the helicity angle ($\theta_{\rm hel}$) between the kaon coming from the $K^{*}$ decay and $B$ meson in the $K\pi$ rest frame is expected to follow a $\rm sin^{2}\theta_{\rm hel}$ distribution for correctly reconstructed $K^{*}$ candidates. On the other hand, misreconstructed candidates follow an asymmetric distribution, with a large number of events migrated towards the region $\cos\theta_{\rm hel} \approx \pm1$. To reject misreconstructed candidates, we require $-0.9 < \cos\theta_{\rm hel} < 0.75$. 

A $K^{*}$ is combined with a prompt photon to reconstruct a $B$ meson. We apply selection on the kinematic variables $M_{\rm bc} = \sqrt{E^{*2}_{\rm beam} - p^{*2}_{B}}$ and $\Delta E = E^{*}_{B} - E^{*}_{\rm beam} $, namely $5.2 < M_{\rm bc} < 5.29$ $\rm GeV\!/c^{2}$ and $-0.4< \rm \Delta E < 0.3$ $\rm GeV$ to suppress combinatorial background. Here $E^{*}_{\rm beam}$ is the beam energy, and $p^{*}_{B}$ and $E^{*}_{B}$ are the momentum and energy of the $B$ meson. The beam energy is tuned to produce a pair of $B$ candidates in an event. Hence, for correctly reconstructed $B$ candidates, we expect the $M_{\rm bc}$ distribution to peak at the nominal $B$ meson mass and the $\Delta E$ distribution to peak at zero. The signal window for $M_{\rm bc}$ and $\Delta E$ variables is defined as, $M_{\rm bc}>5.27 \gevcc$ and $-0.15< \rm \Delta E < 0.07$ $\rm GeV$, which corresponds to around $\pm 3\sigma$ interval. The difference in positive and negative values of $\Delta E$ signal window are due to asymmetry in the $\Delta E$ distribution, which is caused by the energy leakage of high energy photons in the ECL. All the selection criteria are optimized with a figure of merit (FOM) defined as $S/\sqrt{S+B}$, where $S$ and $B$ are the number of signal and background events inside the signal region.

\section{Background suppression}
The dominant background is from $e^{+}e^{-}\to q\overline{q}$ continuum events. The masses of quarks in continuum events are significantly small compared to the ${B}$ meson, hence the former events are highly boosted in the CM frame, which leads to a jet-like topology. On the other hand, the $B$ meson pair is produced almost at rest in the CM frame with decay products having a spherical topology.

To suppress continuum background, a multivariate analyzer~(MVA), namely FastBDT~\cite{r13}, is trained separately for each $K^{*}$ mode using event shape variables. For training and testing the MVA, we have used two independent MC datasets having equal number of correctly reconstructed signal and continuum background events. The signal events are taken from signal MC and continuum background from a 3 $\text{ab}^{-1}$ $q\overline{q}$ MC sample. Half of each sample is used for training and the other half for testing the MVA. The MVA is trained using a total of 17 discriminating variables for neutral modes and 19 for charged modes. These variables include modified Fox-Wolfram moments~\cite{fox_wolfram}, the magnitude of the signal $B$ thrust, the output from the $B$-flavor tagger~\cite{flavor}, the cosine of the angle between the thrust axis of reconstructed $B$ and the thrust axis of the rest-of-event (ROE), the cosine of angle between the thrust axis of reconstructed $B$ and the beam axis, and the cosine of the polar angle between momentum of reconstructed $B$ and the beam axis. A brief description of these discriminating variables can be found in Ref.~\cite{Belle_II_physics}. For each mode, we apply a selection on the MVA output corresponding to the maxima of the FOM. The MVA rejects around 70--90\% of the background, with a signal loss of 10--21\% depending on mode. 

The $M_{\rm bc}$ distribution after the application of the MVA has a significant peaking-background component from $B\overline{B}$ events. The dominant peaking contribution comes from radiative $B$ meson decays to higher kaonic resonances, such as $B\to K^*(1410)\gamma$. The misreconstructed signal, and events evading the $\pi^{0}/\eta$ veto also contribute at the sub-leading order to the peaking-background component. The $\Delta E$ variable is more sensitive to the mass hypothesis of reconstructed tracks and the number of tracks missed while reconstructing a $B$ candidate compared to $M_{\rm bc}$. Thus, a majority of peaking-background events are well separated from the signal in the $\Delta E$ distribution. Hence, the signal yield extraction is performed by fitting the $\Delta E$ variable inside the $M_{\rm bc}$ signal region $M_{\rm bc}>5.27$ GeV$\!/c^{2}$.

After applying all the selection criteria, sometime we are left with more than one reconstructed $B$ candidate per event. If there are multiple $B$ candidates in an event, we retain the one having $M_{\rm bc}$ value closest to the nominal $B$ meson mass. The candidate multiplicity ranges from 1.005--1.090 and the efficiency to select the correctly reconstructed signal from an event with multiple reconstructed $B$ candidates varies from 64--74\% depending on the mode.

\section{Signal yield extraction}

The signal yield is obtained from an unbinned extended maximum-likelihood fit to the $\Delta E$ variable.
For a dataset of $N$ candidates, the likelihood function can be written as:

$$\mathcal{L} = \frac{e^{-(\sum n_{j})}}{N!} \prod^{i=N}_{i=1} \sum^{}_{j} n_{j}P_{j}(\Delta E_{i}),$$ where $P_{j}$ is the probability density function (PDF) and $n_{j}$ is the number of events corresponding to the $j^{th}$ component, respectively. The argument $\Delta E_{i}$ denotes the value of $\Delta E$ for the $i^{th}$ candidate. 
The fit employs in total three components, one each for correctly reconstructed signal, background, and misreconstructed signal events, respectively. We obtain the PDF for correctly reconstructed and misreconstructed signal components by fitting to $\Delta E$ distributions of signal MC events. The PDF for correctly reconstructed signal events is modeled with the sum of a Cruijff function~\cite{RooCruijiff} and a Gaussian. To model the distribution of misreconstructed signal events, we use a Cruijff function. The ratio of misreconstructed to correctly reconstructed signal events is kept fixed to the value obtained from signal MC. The fraction of misreconstructed signal varies from 2--11\% depending on mode. The shape of the background PDF is determined by fitting the $\Delta E$ distribution of events from $B\overline{B}$ and $q\overline{q}$ background MC. The background can be classified into two categories, namely combinatorial and peaking. We model the peaking component with a Gaussian and the combinatorial background using a Chebyshev polynomial. The yields of correctly reconstructed signal, combinatorial, and peaking background events are determined from the fit. 

\begin{figure}[h]
    \subfigure[\hspace{0.1cm} ${B^{0}\to K^{*0} [K^{+} \pi^{-}] \gamma}$ ]	
	{
	\includegraphics[scale=0.15]{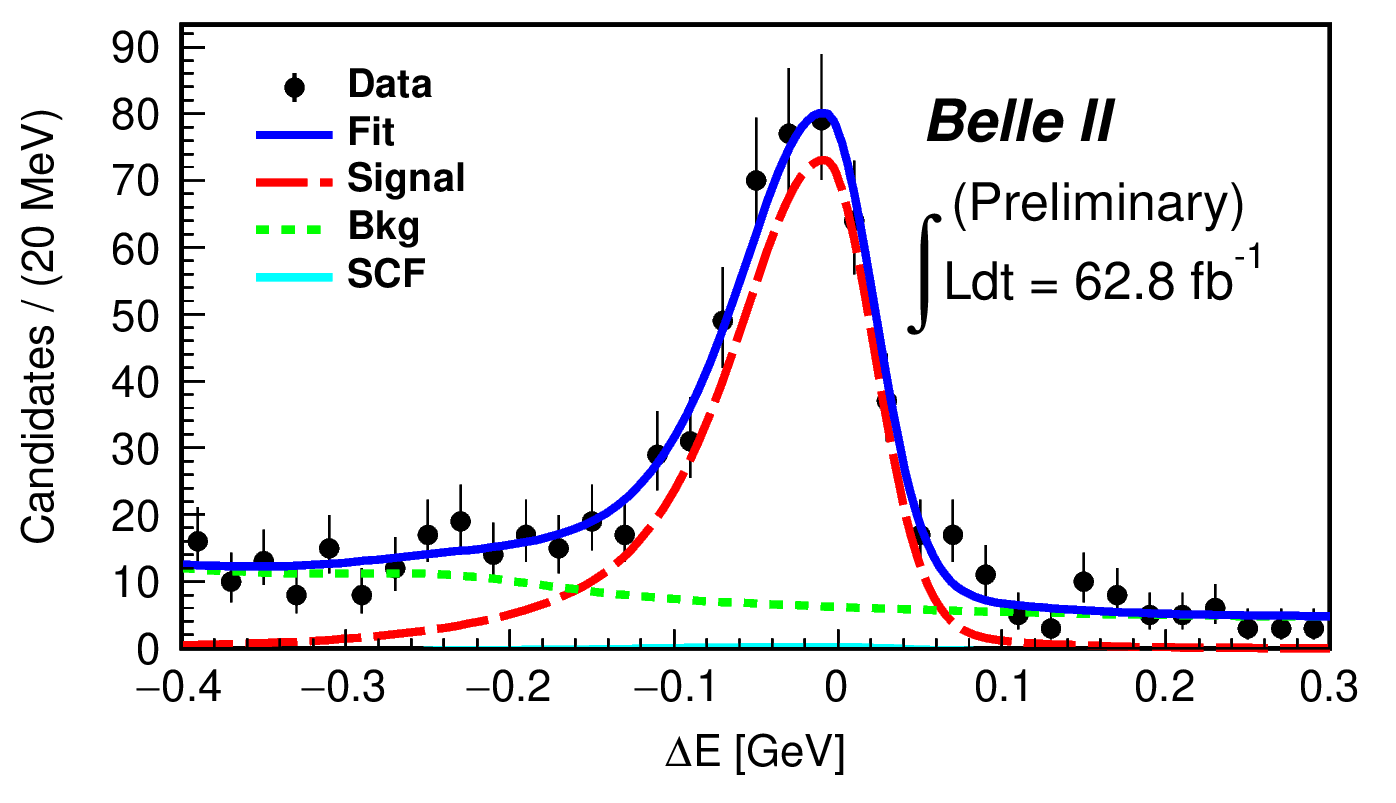}
	}
	\subfigure[\hspace{0.1cm} ${B^{0}\to K^{*0} [K^0_{\rm S} \pi^{0}] \gamma}$ ]
    {
	\includegraphics[scale=0.15]{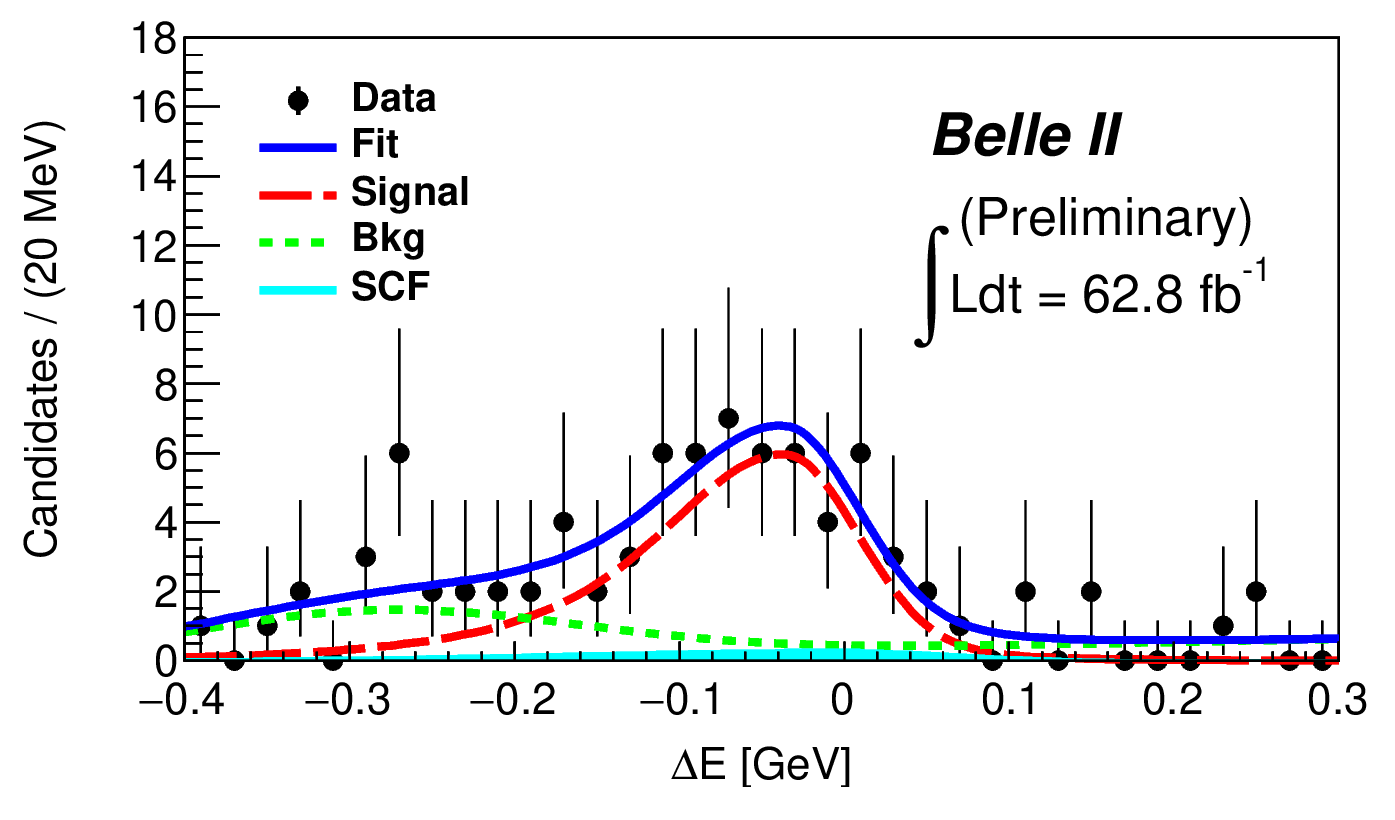}
	}
	\subfigure[\hspace{0.1cm} ${B^{+}\to K^{*+} [K^{+} \pi^{0}] \gamma}$ ]	
	{
	\includegraphics[scale=0.15]{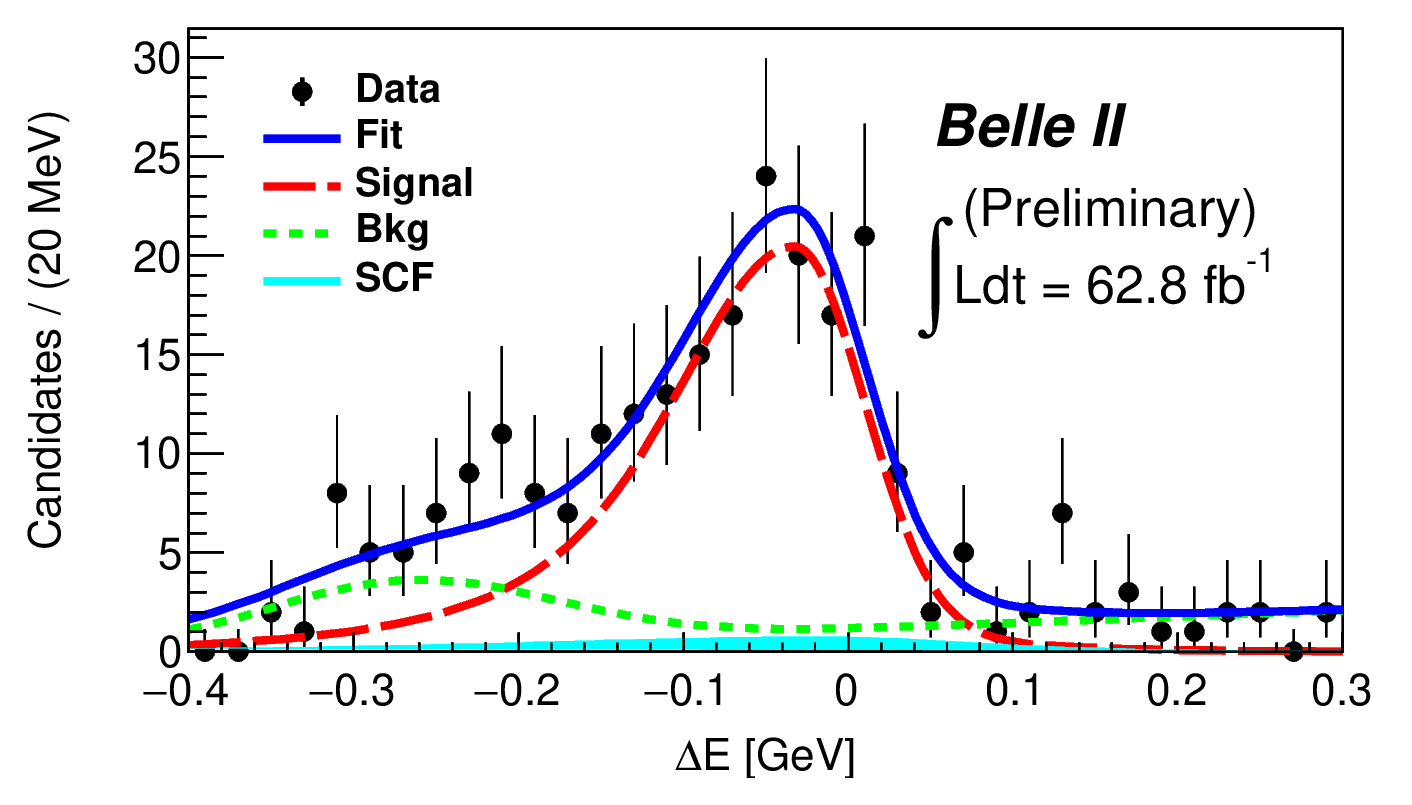}
	}
	\subfigure[\hspace{0.1cm} ${B^{+}\to K^{*+} [K^0_{\rm S} \pi^{+}] \gamma}$ ]	
	{
	\includegraphics[scale=0.15]{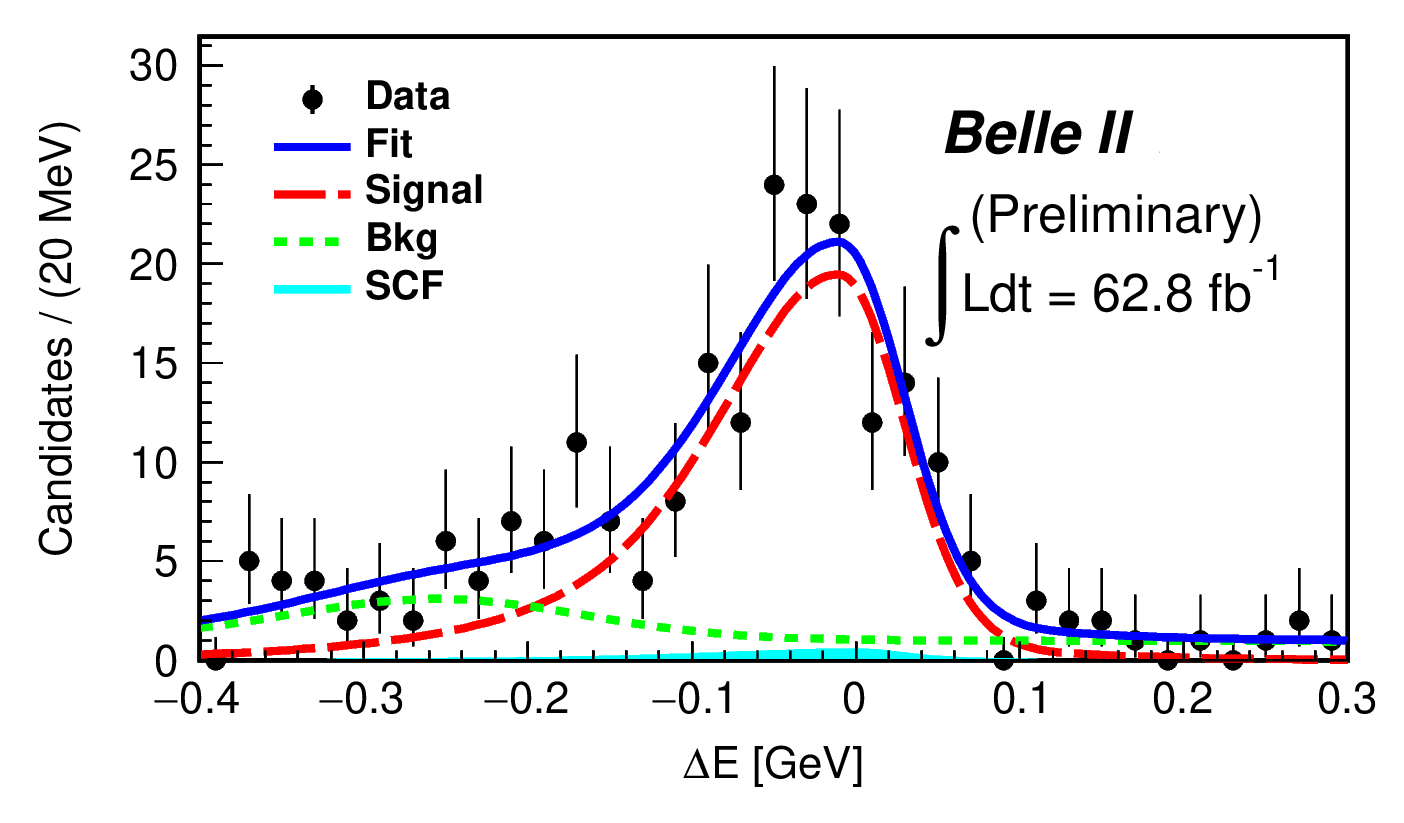}
	}
	\caption{$\Delta E$ distributions for each $B\to K^{*}\gamma$ mode with the fit result superimposed. The black dots with error bars denote the data, the blue curve denotes the total fit, the dashed red curve is the signal component, the dotted green curve is the background component, and the filled cyan region is the misreconstructed signal component.}
	\label{Data_fit}
\end{figure}

An ensemble of 1000 toy datasets is generated using the fit model. These datasets are fitted with the same fit model to study potential fit bias. The pull distributions of fit parameters were consistent with normal distribution within the fit uncertainties, implying that the fit strategy is unbiased. The results of fit performed for all four modes in data are shown in Fig~\ref{Data_fit}.

\FloatBarrier
\section{Measurement of Branching fraction}

The branching fraction is calculated using the following expression:

 $${\mathcal{B}} = \frac{n_{\rm sig}}{2 \times N_{B\overline{B}}\times f^{\pm} (f^{00}) \times \epsilon},$$ where $n_{\rm sig}$ is the signal yield from fit, $\epsilon$ is the signal selection efficiency, $N_{B\overline{B}}$ is the number of $B\overline{B}$ pairs, and $f^{\pm}$ ($f^{00}$) is the branching fraction of $\Upsilon(4\text{S})$ to charged (neutral) $B\overline{B}$ pairs. 
The results are listed in Tables~\ref{table:Fit_Results} and \ref{table:Combined_Results}. The measured branching fractions are compatible with their world average values reported by the Particle Data Group (PDG)~\cite{PDG} at the level of one and two standard deviations for the neutral and charged modes, respectively. Individually none of these deviations is statistically significant, albeit being in the same direction. While the size of the discrepancy of our combined result is only 2.3 standard deviations, the presence of residual systematic effects cannot be ruled out at this stage. Potential sources could be related to the peaking background, which will be systematically investigated in the next iteration of the analysis with more data.

\begin{table}[H]
	\caption{Signal yield, efficiency and measured branching fraction (${\mathcal B}_\mathrm{meas}$) for each mode. When two uncertainties are given, the first is statistical and the second is systematic. The world-average values reported by the PDG are given for comparison.}
	\label{table:Fit_Results}
	\begin{center}
	\begin{tabular}{ |c |c | c| c| c|}
		\hline
		
		Mode & Signal yield & Efficiency (\%) & ${\mathcal B}_\mathrm{meas}$ [$10^{-5}$] & ${\mathcal B}_\mathrm{PDG}$ [$10^{-5}]$ \\
		\hline
		$B^{0} \to K^{*0}[K^{+}\pi^{-}] \gamma$ & $454 \pm 28$ & $15.22\pm 0.03$ & $4.5 \pm 0.3 \pm 0.2$ & $4.18 \pm 0.25$ \\ 
		$B^{0} \to K^{*0}[K^0_{\rm S}\pi^{0}] \gamma$  & $50 \pm 10$ & $1.73\pm 0.01$ & $4.4 \pm 0.9 \pm 0.6$	 & $4.18 \pm 0.25$ \\
		$B^{+} \to K^{*+}[K^{+}\pi^{0}] \gamma$ & $169 \pm 18$ & $4.84\pm 0.02$ & $5.0 \pm 0.5 \pm 0.4$ & $3.92 \pm 0.22$\\
		$B^{+} \to K^{*+}[K^0_{\rm S}\pi^{+}] \gamma$ & $160 \pm 17$ & $4.23\pm0.02$ & $5.4 \pm 0.6 \pm 0.4$ & $3.92 \pm 0.22$\\ 
		\hline
		
	\end{tabular}
	\end{center}
\end{table}

\begin{table}[H]
	\caption{Measured branching fraction (${\mathcal B}_\mathrm{meas}$) for combined charged and neutral modes. The first uncertainty is statistical and the second is systematic. The world-average values reported by the PDG are given for comparison.}
	\label{table:Combined_Results}
	\begin{center}
	\begin{tabular}{ |c | c| c|}
		\hline
		
		Mode & ${\mathcal B}_\mathrm{meas}$ [$10^{-5}$] & ${\mathcal B}_\mathrm{PDG}$ [$10^{-5}$] \\
		\hline
		$B^{0} \to K^{*0}\gamma$ & $4.5 \pm 0.3 \pm 0.2$ & $4.18 \pm 0.25$ \\ 
		$B^{+} \to K^{*+}\gamma$ & $5.2 \pm 0.4 \pm 0.3$ & $3.92 \pm 0.22$\\
		\hline
		
	\end{tabular}
	\end{center}
\end{table}

\section{Systematic Uncertainties}

In this section, we describe the various sources of systematic uncertainties. A systematic uncertainty of 1.6\% is assigned to the uncertainty in the number of $B\overline{B}$ events~\cite{nbb}. The performance of $\pi^{0}/\eta$ veto between data and simulation is studied using $B^{+}\to D^{0}[\to K^{+}\pi^{-}]\pi^{+}$ and $B^{0}\to D^{0}[\to K^{+}\pi^{-}\pi^{-}]\pi^{+}$ samples, where the fast pion coming from the $B$ decay is treated as a photon candidate. We assign a systematic uncertainty of 3.8\% due to application of $\pi^{0}/\eta$ veto. The uncertainty in the selection efficiency of high energy photon candidates is estimated from a control sample of radiative dimuon events. The difference in efficiency between data and simulation due to PID selection for charged hadrons is calculated using a $D^{*+} \to D^{0}[K^{-}\pi^{+}]\pi^{+}$ control sample~\cite{kaons}. The corrections are calculated in bins of the momentum and cosine of the polar angle. We assign a systematic uncertainty of 0.6\% for pion and 0.8\% for kaon selection. Comparing the reconstruction efficiency of $K^0_{\rm S}$ between data and simulation, a systematic uncertainty of 2.4\% is assigned. The difference in reconstruction efficiency of $\pi^{0}$ between data and simulation is studied by comparing the yield of $\eta$ meson between the $\eta \to \gamma\gamma$ and $\eta \to \pi^{0}\pi^{0}\pi^{0}$ channels. A systematic uncertainty of 3.4\% is assigned for $\pi^{0}$ selection. We assign a systematic uncertainty of 0.7\% per charged track~\cite{tracking}, which results in a systematic uncertainty of 0.7\% for $K^{*+}[K^{+}\pi^{0}]$ and 1.4\% for rest of the modes. The systematic uncertainty due to MVA criteria are studied in the off-resonance sideband. We assign a systematic uncertainty of 2--6\% depending on mode. The uncertainty due to limited statistics of signal MC is 0.2--0.5\% depending on mode. Various PDF shape parameters fixed while performing the fit are varied by $\pm 1\sigma$ around their mean values. The obtained variation in the signal yield is taken as a systematic uncertainty. The misidentification of tracks coming from signal events gives rise to misreconstructed signal candidates. The fraction of such misreconstructed candidates is fixed to the values obtained from simulation while performing fit to data. The fraction of misreconstructed signal events is varied by $\pm 100\%$ around its nominal value to assign a systematic uncertainty.

We summarize the systematic uncertainties for the branching fraction measurements in Table~\ref{table:Systematic}. The individual sources of uncertainties are assumed to be independent and are added in quadrature to arrive at the total uncertainty.

\begin{table}[H]

    \caption{ Relative systematic uncertainties (in \%) for the branching fraction measurement. }
	\label{table:Systematic}
	\begin{center}
		\begin{tabular}{  | l | c| c| c| c|}
		\hline

        Source &$K^{*0}[K^{+}\pi^{-}]\gamma$ & $K^{*0}[K^0_{\rm S}\pi^{0}]\gamma$ & $K^{*+}[K^{+}\pi^{0}]\gamma$ & $K^{*+}[K^0_{\rm S}\pi^{+}]\gamma$ \\

        \hline   
        No. of $B\overline{B}$ events   & 1.6 & 1.6 & 1.6 & 1.6\\
        Photon selection & $^{+0.2}_{-0.4}$ & $^{+0.2}_{-0.4}$ & $^{+0.2}_{-0.4}$ & $^{+0.2}_{-0.4}$ \\
        $\pi^{0}/\eta$ veto & 3.8 & 3.8 & 3.8 & 3.8 \\
        Pion identification & 0.6 & --- & --- & 0.6 \\
        Kaon identification  & 0.8 & --- & 0.8 & --- \\
        $K^0_{\rm S}$ reconstruction  & --- & 2.4 & --- & 2.4  \\
        $\pi^{0}$ selection & --- & 3.4 & 3.4 & --- \\
        Tracking efficiency & 1.4 & 1.4 & 0.7 & 1.4 \\
        MVA selection & 2.0 & 6.0 & 2.0 & 4.0 \\
        MC statistics  & 0.2 & 0.5 & 0.3 & 0.3 \\
        PDF shape parameters & 1.0 & $^{+7.4}_{-5.4}$ & $^{+2.4}_{-3.1}$  & $^{+0.6}_{-1.4}$ \\
        Misreconstructed signal & 1.5 & $^{+6.8}_{-7.2}$ & $^{+4.7}_{-5.9}$ & $^{+2.5}_{-3.1}$ \\
        \hline
        Total & 5.3 & $^{+13.2}_{-12.4}$ & $^{+7.9}_{-8.9}$ & $^{+7.0}_{-7.3}$ \\
        \hline
		\end{tabular}
	\end{center}
\end{table}

\section{Summary and Conclusion}
We report the measurements by the Belle II experiment of $B\to K^{*}\gamma$ branching fractions. The  measured values of the branching fractions are consistent with the world average values at the level of one and two standard deviations for neutral and charged modes, respectively. As Belle II collects more data, we will report the results for measurement of the observables $A_{CP}$ and $\Delta_{0+}$ along with the branching fractions.

\section{Acknowledgement}
We thank the SuperKEKB group for the excellent operation of the
accelerator; the KEK cryogenics group for the efficient
operation of the solenoid; the KEK computer group for
on-site computing support; and the raw-data centers at
BNL, DESY, GridKa, IN2P3, and INFN for off-site computing support.
This work was supported by the following funding sources:
%Armenia
Science Committee of the Republic of Armenia Grant No. 20TTCG-1C010;
%Australia
Australian Research Council and research grant Nos.
DP180102629, 
DP170102389, 
DP170102204, 
DP150103061, 
FT130100303, 
FT130100018,
and
FT120100745;
%Austria
Austrian Federal Ministry of Education, Science and Research,
Austrian Science Fund No. P 31361-N36, and
Horizon 2020 ERC Starting Grant no. 947006 ``InterLeptons''; 
%Canada
Natural Sciences and Engineering Research Council of Canada, Compute Canada and CANARIE;
%China
Chinese Academy of Sciences and research grant No. QYZDJ-SSW-SLH011,
National Natural Science Foundation of China and research grant Nos.
11521505,
11575017,
11675166,
11761141009,
11705209,
and
11975076,
LiaoNing Revitalization Talents Program under contract No. XLYC1807135,
Shanghai Municipal Science and Technology Committee under contract No. 19ZR1403000,
Shanghai Pujiang Program under Grant No. 18PJ1401000,
and the CAS Center for Excellence in Particle Physics (CCEPP);
%Czech Republic
the Ministry of Education, Youth and Sports of the Czech Republic under Contract No.~LTT17020 and 
Charles University grants SVV 260448 and GAUK 404316;
%EU
European Research Council, 7th Framework PIEF-GA-2013-622527, 
Horizon 2020 ERC-Advanced Grants No. 267104 and 884719,
Horizon 2020 ERC-Consolidator Grant No. 819127,
Horizon 2020 Marie Sklodowska-Curie grant agreement No. 700525 `NIOBE,' 
and
Horizon 2020 Marie Sklodowska-Curie RISE project JENNIFER2 grant agreement No. 822070 (European grants);
%France
L'Institut National de Physique Nucl\'{e}aire et de Physique des Particules (IN2P3) du CNRS (France);
%Germany
BMBF, DFG, HGF, MPG, and AvH Foundation (Germany);
%India
Department of Atomic Energy under Project Identification No. RTI 4002 and Department of Science and Technology (India);
%Israel
Israel Science Foundation grant No. 2476/17,
United States-Israel Binational Science Foundation grant No. 2016113, and
Israel Ministry of Science grant No. 3-16543;
%Italy
Istituto Nazionale di Fisica Nucleare and the research grants BELLE2;
%Japan
Japan Society for the Promotion of Science,  Grant-in-Aid for Scientific Research grant Nos.
16H03968, 
16H03993, 
16H06492,
16K05323, 
17H01133, 
17H05405, 
18K03621, 
18H03710, 
18H05226,
19H00682, % Niigata
26220706,
and
26400255,
the National Institute of Informatics, and Science Information NETwork 5 (SINET5), 
and
the Ministry of Education, Culture, Sports, Science, and Technology (MEXT) of Japan;  
%Korea
National Research Foundation (NRF) of Korea Grant Nos.
2016R1\-D1A1B\-01010135,
2016R1\-D1A1B\-02012900,
2018R1\-A2B\-3003643,
2018R1\-A6A1A\-06024970,
2018R1\-D1A1B\-07047294,
2019K1\-A3A7A\-09033840,
and
2019R1\-I1A3A\-01058933,
Radiation Science Research Institute,
Foreign Large-size Research Facility Application Supporting project,
the Global Science Experimental Data Hub Center of the Korea Institute of Science and Technology Information
and
KREONET/GLORIAD;
%Malaysia
Universiti Malaya RU grant, Akademi Sains Malaysia and Ministry of Education Malaysia;
%Mexico
% CINVESTAV-IPN, UNAM, UAS, BUAP and CONACYT are funded under
Frontiers of Science Program contracts
FOINS-296,
CB-221329,
CB-236394,
CB-254409,
and
CB-180023, and SEP-CINVESTAV research grant 237 (Mexico);
%Poland
the Polish Ministry of Science and Higher Education and the National Science Center;
%Russia
the Ministry of Science and Higher Education of the Russian Federation,
Agreement 14.W03.31.0026, and
the HSE University Basic Research Program, Moscow;
%Saudi Arabia
University of Tabuk research grants
S-0256-1438 and S-0280-1439 (Saudi Arabia);
%Slovenia
Slovenian Research Agency and research grant Nos.
J1-9124
and
P1-0135; 
%Spain
Agencia Estatal de Investigacion, Spain grant Nos.
FPA2014-55613-P
and
FPA2017-84445-P,
and
CIDEGENT/2018/020 of Generalitat Valenciana;
%Taiwan
Ministry of Science and Technology and research grant Nos.
MOST106-2112-M-002-005-MY3
and
MOST107-2119-M-002-035-MY3, 
and the Ministry of Education (Taiwan);
%Thailand
Thailand Center of Excellence in Physics;
%Turkey
TUBITAK ULAKBIM (Turkey);
%Ukraine
Ministry of Education and Science of Ukraine;
%USA
the US National Science Foundation and research grant Nos.
PHY-1807007 % Luther
and
PHY-1913789, % Indiana CEEM
and the US Department of Energy and research grant Nos.
DE-AC06-76RLO1830, % PNNL
DE-SC0007983, % Wayne State
DE-SC0009824, % Florida
DE-SC0009973, % VPI
DE-SC0010073, % South Carolina
DE-SC0010118, % Carnegie Mellon
DE-SC0010504, % Hawaii
DE-SC0011784, % Cincinnati
DE-SC0012704, % BNL
DE-SC0021274; % Mississippi
%last group
and
%Vietnam
the Vietnam Academy of Science and Technology (VAST) under grant DL0000.05/21-23.


\begin{references}
	%\section{References}
	\bibitem{CKM}
	 M. Kobayashi and T. Maskawa, \href{https://academic.oup.com/ptp/article/49/2/652/1858101}{Prog. Theor. Phys. \textbf{49}, 652 (1973)}.
	 
	 \bibitem{BSM1}
	 M. Benzke et al., \href{https://journals.aps.org/prl/abstract/10.1103/PhysRevLett.106.141801}{Phys. Rev. Lett. \textbf{106}, 141801 (2011)}.
	 
	 \bibitem{BSM2}
	J. Lyon and R. Zwicky, Isospin asymmetries in $B\to (K^{*},\rho)\gamma/\ell^{+}\ell^{-}l$ and $B\to K\ell^{+}\ell^{-}l$ in and beyond the Standard Model, \href{https://journals.aps.org/prd/abstract/10.1103/PhysRevD.88.094004}{Phys. Rev. D \textbf{88}, 094004 (2013)}.
	 
	\bibitem{SM1}
	A. Ali, B. Pacjak, and C. Greub, Towards $B\to V \gamma$ Decays at NNLO in SCET, \href{https://link.springer.com/article/10.1140%2Fepjc%2Fs10052-008-0623-5}{Eur. Phys. J. C \textbf{55} 577-595 (2008)}.
	
	\bibitem{SM2}
	A. Paul and D.~M. Straub, Constraints on new physics from radiative B decays. \href{https://link.springer.com/article/10.1007%2FJHEP04%282017%29027}{J. High Energ. Phys. 2017, \textbf{27} (2017)}.
	
	\bibitem{SM3}
	M. Matsumori, A. I. Sanda, and Y.-Y. Keum, \href{https://journals.aps.org/prd/abstract/10.1103/PhysRevD.72.014013}{Phys. Rev. D \textbf{72}, 014013 (2005)}.
	
	\bibitem{SM4}
	W. Altmannshofer and D.~M. Straub, New physics in $b\to s$ transitions after LHC run 1. \href{https://link.springer.com/article/10.1140/epjc/s10052-015-3602-7}{Eur. Phys. J. C \textbf{75}, 382 (2015)}.
	
	\bibitem{Belle_paper}
	T. Horiguchi et al. (Belle Collaboration), \href{https://journals.aps.org/prl/abstract/10.1103/PhysRevLett.119.191802}{Phys. Rev. Lett. \textbf{119}, 191802 (2017)}.
	
	\bibitem{CLEO_paper}
	Coan et al. (CLEO Collaboration), \href{https://journals.aps.org/prl/abstract/10.1103/PhysRevLett.84.5283}{Phys. Rev. Lett. \textbf{84} (2000)}.
	
	\bibitem{BaBar_paper}
	B. Aubert et al. (BaBar Collaboration), \href{https://journals.aps.org/prl/abstract/10.1103/PhysRevLett.103.211802}{Phys. Rev. Lett. \textbf{103}, 211802 (2009)}.
	
	\bibitem{superkekb} K. Akai et al. (SuperKEKB Accelerator Team), SuperKEKB Collider, \href{https://reader.elsevier.com/reader/sd/pii/S0168900218309616?token=30564431B53C974D2AD29B8A4CE3AA04536A423B0CE6B92417A504BCCF2BC478ECB05D003A00DAFF4072825650BA0373&originRegion=eu-west-1&originCreation=20210805174928}{Nucl. Instrum. Meth. A \textbf{907} (2018)}.
	
	\bibitem{belle2tdr} T. Abe et al. (Belle II Collaboration), Belle II Technical Design Report, (2010), \href{https://arxiv.org/abs/1011.0352}{arXiv:1011.0352}.
	
	\bibitem{nbb} F.~Abudin{\'e}n et al. (Belle II Collaboration), \href{https://docs.belle2.org/record/1997}{BELLE2-NOTE-PL-2020-006 (2020)}.

    \bibitem{EVETGEN} D. Lange, \href{https://www.sciencedirect.com/science/article/abs/pii/S0168900201000894}{Nucl. Instrum. Meth. A \textbf{462}, 152 (2001)}.

    \bibitem{GEANT4} S. Agostinelli et al., \href{https://www.sciencedirect.com/science/article/abs/pii/S0168900203013688?via\%3Dihub}{Nucl. Instrum. Meth. A \textbf{506}, 250 (2003).}
	
	\bibitem{basf2} T. Kuhr et al., The Belle II Core Software, \href{https://link.springer.com/article/10.1007/s41781-018-0017-9}{Comput. Softw. Big Sci. \textbf{3} (2019) no. 1, 1}.
	
	\bibitem{PSDMVA} S. Longo et al., CsI(Tl) pulse shape discrimination with the Belle II electromagnetic calorimeter as a novel method to improve particle identification at electron–positron colliders, \href{https://www.sciencedirect.com/science/article/abs/pii/S0168900220309591?via%3Dihub}{Nucl. Instrum. Meth. A \textbf{982}, (2020) 164562}.
	
	\bibitem{r13} T. Keck, FastBDT: A speed-optimized and cache-friendly implementation of stochastic gradient-boosted decision trees for multivariate classification, \href{https://arxiv.org/abs/1609.06119}{arXiv:1609.06119}.
	
	\bibitem{fox_wolfram}
	G.~C. Fox and S. Wolfram, \href{https://journals.aps.org/prl/abstract/10.1103/PhysRevLett.41.1581}{Phys. Rev. Lett. \textbf{41}, 1581 (1978)}.

	\bibitem{flavor}
	F.~Abudin{\'e}n, Ph.D. Thesis, Development of a $B^{0}$ flavor tagger and performance study of a novel time-dependent CP analysis of the decay $B^{0}\to \pi^{0}\pi^{0}$ at Belle II, Ludwig Maximilian University of Munich (2018), \href{https://docs.belle2.org/record/1215?ln=en,}{BELLE2-PTHESIS-2018-003}.
	
	\bibitem{Belle_II_physics}
	E. Kou et al. (Belle II Collaboration), \href{https://academic.oup.com/ptep/article/2019/12/123C01/5685006}{Prog. Theor. Exp. Phys. \textbf{2019}, 123C01 (2019)}.
	
	\bibitem{RooCruijiff}
    P.  del  Amo Sanchez et al. (BaBar  Collaboration), \href{https://journals.aps.org/prd/abstract/10.1103/PhysRevD.82.051101}{Phys. Rev. D \textbf{82} (2010) 051101}. 	

    \bibitem{PDG} P.~A. Zyla et al. (Particle Data Group), \href{https://academic.oup.com/ptep/article/2020/8/083C01/5891211}{Prog. Theor. Exp. Phys. \textbf{2020}, 083C01 (2020)}.


    \bibitem{kaons} F.~Abudin{\'e}n et al. (Belle II Collaboration), \href{https://docs.belle2.org/record/2052}{BELLE2-NOTE-PL-2020-024 (2020)}.
	

    \bibitem{tracking} F.~Abudin{\'e}n et al. (Belle II Collaboration), \href{https://docs.belle2.org/record/2035}{BELLE2-NOTE-PL-2020-014 (2020)}.


	
	\end{references}
\end{document}